# Advances in optical coherence tomography for dermatology

Olivier Levecq[a], Arthur Davis[a,b], Hicham Azimani[a], David Siret[a], Jean-Luc Perrot[c], Arnaud Dubois*[a,b]

[a]DAMAE Medical, 28 rue de Turbigo, 75003 Paris, France;
[b]Laboratoire Charles Fabry, Institut d'Optique Graduate School, Université Paris-Sud, 91127 Palaiseau Cedex, France
[c]Service dermatologie, CHU St-Etienne, 42055 Saint-Etienne, France

**ABSTRACT**

This paper reports on advances in optical coherence tomography (OCT) for application in dermatology. Full-field OCT is a particular approach of OCT based on white-light interference microscopy. FF-OCT produces *en face* tomographic images by arithmetic combination of interferometric images acquired with an area camera and by illuminating the whole field of view with low-coherence light. The major interest of FF-OCT lies in its high imaging spatial resolution (~ 1.0 µm) in both lateral and axial directions, using a simple and robust experimental arrangement. Line-field OCT (LF-OCT) is a recent evolution of FF-OCT with line illumination and line detection using a broadband spatially coherent light source and a line-scan camera in an interference microscope. LF-OCT and FF-OCT are similar in terms of spatial resolution. LF-OCT has a significant advantage over FF-OCT in terms of imaging penetration depth due to the confocal gate achieved by line illumination and detection. B-scan imaging using FF-OCT requires the acquisition of a stack of *en face* images, which usually prevents *in vivo* applications. B-scan imaging using LF-OCT can be considerably faster due to the possibility of using a spatially coherent light source with much higher brightness along with a high-speed line camera. Applied in the field of dermatology, the LF-OCT images reveal a comprehensive morphological mapping of skin tissues *in vivo* at a cellular level similar to histological images.

## 1. INTRODUCTION

Skin cancer is the most common form of all human cancers[1,2]. Recent studies show the number of skin cancer cases growing at an alarming rate[3]. In addition to causing illness and death, skin cancer is a huge economic burden. Early detection is likely the most promising way to reduce morbidity and mortality. The standard of care procedure for the detection of skin cancer involves visual examination of the superficial structures of the skin using a dermatoscope[4-6]. The clinical and dermoscopic evaluation of skin lesions is then followed by a biopsy, which is a surgical removal of a tissue sample in order to examine it at the cellular-level using an optical microscope. In this procedure, nearly 60% of all skin biopsies result in benign diagnoses[2]. However, approximately 20% of skin cancers, including roughly a third of all melanomas, are missed[3]. Given this challenge, improved diagnostic modalities using noninvasive imaging techniques have been developed to provide earlier, more accurate detection of suspicious lesions, thereby decreasing the false positive rates of dermoscopy and thereby the rate of unnecessary biopsies[7]. The clinically available techniques capable of *in vivo* skin imaging with the highest spatial resolution are reflectance confocal microscopy, nonlinear optical microscopy, and optical coherence tomography[8,9].

Reflectance confocal microscopy (RCM) is an optical imaging technique that allows the analysis of the skin with a nearly histological resolution (~ 1 µm), *i.e.* at a cellular level[10-13]. Similar to dermoscopy images, real-time images obtained by RCM are oriented horizontal to the skin surface (*en face* sections). RCM has been applied in the clinical arena for the diagnosis of melanocytic and nonmelanocytic lesions where it has been proven to increase the diagnostic accuracy when coupled with dermoscopy[10,11]. Beyond its application in skin oncology, RCM can be useful to delineate indications for inflammatory and infectious skin conditions. The main limitation of RCM is its relatively low penetration in skin because of strong light scattering. A penetration depth of ~ 200 µm can be achieved, which is not sufficient to image in the dermis[8]. Also problematic is the fact that RCM sections are oriented perpendicular to conventional histological sections, making them difficult to interpret.

*arnaud.dubois@institutoptique.fr

Nonlinear optical microscopy is a high-resolution imaging modality based on nonlinear interactions of light with biological tissues[14]. Compared with optical coherence tomography (OCT), nonlinear microscopy offers a better spatial resolution, similar to that of RCM. Advances in developing multiphoton excitation microscopes with novel contrast mechanisms, such as second harmonic generation, further allow visualization of skin morphology and function[15]. Key limitations of nonlinear optical microscopy are the orientation of the images (*en face* sections, like RCM), the small field of view (350 μm × 350 μm with the DermaInspect device, Jenlab), and the relatively weak penetration in skin (~ 200 μm).

Optical coherence tomography (OCT) is an interferometric imaging modality initially introduced in the clinical field of ophthalmology[16]. The first use of OCT in dermatology was demonstrated in 1997[17] and since several manufacturers of OCT systems have made OCT imaging for dermatology purposes commercially available[18]. OCT can create *in vivo* cross-sectional and/or *en face* images of skin with an axial resolution of 3-5 μm and a lateral resolution of 3-8 μm[18]. The penetration depth of OCT is ~ 1 mm. The possibility to evaluate OCT images in a cross-sectional view (perpendicular to the skin surface) makes them easier to compare with histology sections. Compared with RCM and nonlinear optical microscopy, OCT has a larger lateral field of view, ranging from 2 to 10 mm. The biggest potential of OCT in dermatology has thus far been in the diagnosis, delineation, and treatment of non-melanoma skin cancers, especially basal cell carcinomas. Pigmented lesions, on the other hand, continue to pose great challenges in OCT imaging, mainly because of an insufficient imaging spatial resolution[19-22].

## 2. OPTICAL COHERENCE TOMOGRAPHY

### 2.1 Time-domain and frequency-domain OCT

Two main categories of OCT can be distinguished: time-domain OCT (TD-OCT) and frequency-domain OCT (FD-OCT)[23]. In TD-OCT, the sample reflectivity profile as a function of depth is acquired point-by-point by scanning the depth in the sample (A-scan). In FD-OCT, the sample reflectivity profile as a function of depth is acquired without scanning the depth, by measuring the spectrum of the interferometric signal. In both TD-OCT and FD-OCT, a B-scan image is then obtained by lateral scanning of the light beam to acquire several adjacent A-scans.

The axial (depth) resolution in OCT is essentially governed by the temporal coherence of the detected light. Improvement of the axial resolution in OCT has been obtained by the emergence of efficient broadband light sources. Outstanding axial resolutions of ~1 μm have been achieved in particular with mode-locked lasers[3] and more recently with supercontinuum laser sources[24,25].

FD-OCT is superior to TD-OCT in terms of acquisition rate and detection sensitivity[27], but presents shortcomings, including a limited lateral resolution[23]. Because all points along the depth range in the sample need to be in focus simultaneously, a depth of focus (DOF) at least equal to the depth range is needed, which limits the beam focusing. Several approaches have been reported to improve the lateral resolution in FD-OCT, including extending the DOF by illuminating the sample with a Bessel beam[28-30] or by using appropriate phase masks[31]. Computational imaging solutions have also been proposed, including interferometric synthetic aperture microscopy[32] and digital refocusing[33,34]. An alternative approach consists of combining several B-scan images acquired at different depths over a reduced DOF with higher lateral resolution. This approach has been implemented using the Gabor-based fusion method[35]. It can also be applied by focusing several light beams at different depths[36]. The latter approach, implemented in the Vivosight FD-OCT device (Michelson Diagnostics) for skin imaging, yields a lateral resolution of 7 μm[36].

Unlike FD-OCT, TD-OCT offers the possibility of adjusting the focus as a function of depth, which makes TD-OCT more appropriate to produce high-resolution images. Dynamic focus tracking in TD-OCT with free-space optics has been reported, but the tracking rate was slow[37]. A microelectromechanical systems (MEMS) mirror was designed for high-speed dynamic focus tracking, but without demonstration of the imaging capability *in vivo*[38]. Another approach consists of acquiring a sequence of images by gradually shifting the focus onto the sample and then fusing together the in-focus imaging zones[24]. This process results in a trade-off between lateral resolution and image acquisition speed. Another method is to collect multiple foci simultaneously with a multifocus fiber tip array[39]. Despite these advances, however, TD-OCT imaging at high lateral resolution using dynamic focusing remains challenging since a high tracking speed is required.

## 2.2 Full-field OCT

Full-field optical coherence tomography (FF-OCT), also often referred to as full-field optical coherence microscopy (FF-OCM), was introduced a few years ago as an alternative to conventional TD-OCT[40-42]. FF-OCT uses an interference microscope and an area camera combined with a low coherence illumination source. *En face* tomographic images are obtained without lateral scanning. FF-OCT has been applied to high resolution (reaching ~ 1.0μm at ~ 700 nm center wavelength) non-invasive three-dimensional imaging of various semi-transparent samples[43,44].

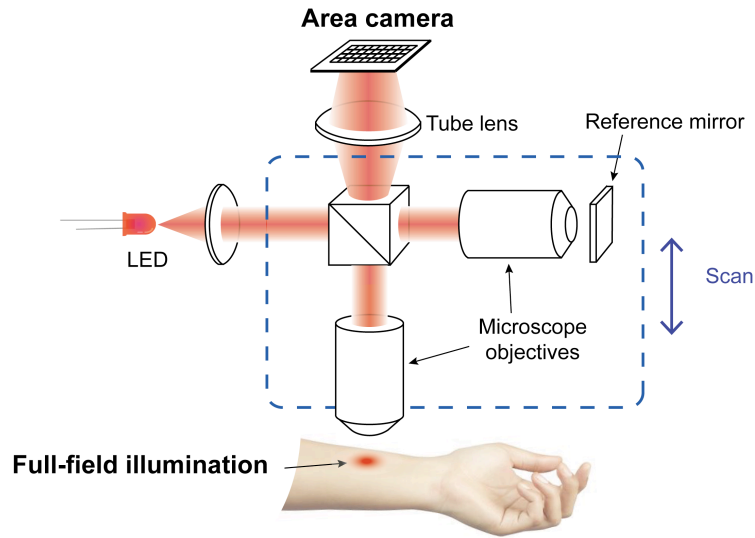

Figure 1. Experimental setup of full-field OCT, based on white-light interference microscopy. The prototype is based on a Michelson interferometer with a microscope objective in each arm (Linnik-type interference microscope). The skin tissue is full-field illuminated using a broadband light emitting diode (LED). The interferometric signal is detected by an area camera. The whole interferometer is displaced to scan the depth in the tissue. Tomographic images of the backscattering skin microstructures are obtained by digital extraction of the interference fringe envelope.

The experimental arrangement of conventional FF-OCT is based on a white-light interference microscope. Several configurations of interference microscopes can be used in FF-OCT. The most common configuration is the Linnik configuration[40] because its offers high flexibility and control. A scheme of the typical FF-OCT experimental setup is shown in Figure 1. It consists of a Michelson-type interferometer with a microscope objective placed in each arm. A halogen lamp or a broadband light emitting diode (LED) provides uniform illumination of the microscope objective fields with low coherence light. A low reflectivity mirror (reflectivity < 5 %) is placed in the reference arm of the interferometer in the focal plane of the microscope objective. An area camera acquires the interferometric images delivered by the microscope. Immersion microscope objectives are employed to minimize the mismatch between the coherence plane and the objective focal plane and to minimize the mismatch of optical dispersion in the interferometer arms, as the imaging depth in the sample increases[45]. The use of an immersion medium also reduces light reflection at the sample surface. A motorized translation stage is used to translate the sample relatively to the microscope in the axial direction to scan the depth in the sample.

The signal in FF-OCT - the amplitude of the acquired interferometric signal, *i.e.* the fringe envelope - is obtained by phase-shifting interferometry, a well-established method for phase measurements that can also be applied for measuring the amplitude of the interferometric signal[38-40]. The method consists of introducing a time-modulation of the relative phase between the reference and sample waves and the acquiring of several interferometric images. In most FF-OCT systems, *en face* tomographic images are produced by an arithmetic combination of four interferometric images using a four-frame phase-shifting algorithm[42]. The phase-shift is generated either by the displacement of the reference mirror attached to a piezoelectric actuator or by the scan of the depth in the sample. Each tomographic image corresponds to an image of the reflecting/backscattering structures of the sample located in a "slice", perpendicular to the optical axis (*en face* image). The width of this "slice", equal to the temporal coherence length of the detected light, defines the axial imaging resolution. The tomographic images can be produced at a maximum rate equal to a fraction of the camera frame rate (depending on the number of frames required in the phase-shifting algorithm, *i.e.* usually a quarter of the camera frame rate). However, in practice it is necessary to accumulate several interferometric images to improve the detection sensitivity. The *en face* tomographic images are then produced at a frequency on the order of a few Hertz. B-scan imaging using FF-OCT requires the acquisition of a stack of *en face* images, which may take several seconds. B-scan imaging using FF-OCT is usually not applicable to *in vivo* studies.

## 2.3 Line-field OCT

Line-field optical coherence tomography (LF-OCT) is an evolution of FF-OCT with line illumination and line detection. LF-OCT can be seen as a parallelization of TD-OCT. Since multiple A-scans are acquired in parallel using a line camera, only one scan (depth-scan) is required to produce a B-scan image. The scan of the depth can thus be slower compared to conventional TD-OCT, without increasing the image acquisition time, which makes dynamic focusing easier. Moreover, line illumination and detection, combined with the use of microscope objectives with relatively high numerical aperture (NA), provides a confocal gate, which minimizes the amount of unwanted scattered light that may be detected by the camera. In LF-OCT, the amount of light incident onto the photodetector, that does not contribute to the interference is thus significantly reduced compared to FF-OCT[46]. This confers to LF-OCT a significant advantage over FF-OCT in terms of imaging penetration depth. LF-OCT is thus similar to FD-OCT in terms of imaging penetration in scattering tissues such as skin. Moreover, B-scan imaging can be considerably faster with LF-OCT compared to FF-OCT due to the possibility of using a spatially coherent light source with much higher brightness along with a high-speed line camera.

In our LF-OCT prototype, the acquired interferometric data are processed using a Fourier transform method to generate intensity-based tomographic images. The calculations are performed with a field-programmable gate array (FPGA) to optimize the operation speed. B-scan images, after being appropriately rescaled, are displayed in real-time (10 frame/s). By using a supercontinuum laser as the light source and balancing the optical dispersion in the interferometer arms, the axial resolution of LF-OCT images is comparable to the best axial resolution achieved by OCT at ~ 800 nm center wavelength. An isotropic spatial resolution of ~ 1 μm is achieved using microscope objectives of NA = 0.5 (similar to FF-OCT). Applied in the field of dermatology, the LF-OCT images reveal a comprehensive morphological mapping of skin tissues at a cellular level similar to histological images, *in vivo* (see Figure 2).

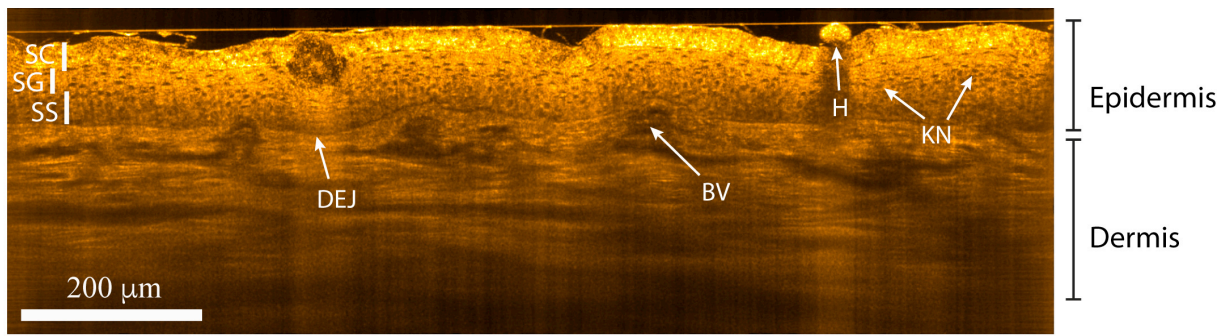

Figure 2. LF-OCT image of healthy human skin (phototype 2) on the back of the hand. SC: stratum corneum layer; SG: stratum granulosum layer with stretch nuclei; SS: stratum spinosum layer with roundish nuclei; BV: blood vessel; KN: keratinocytes nucleus; DEJ: dermal-epidermal junction; H: hair.

## 3. CONCLUSION

LF-OCT provides images that are close to histological images in terms of spatial resolution and orientation. The technique has been developed by the startup company DAMAE Medical, leading in 2016 to OCTAV®, a prototype device for research applications. Images of skin lesions generated by OCTAV® have been enthusiastically received by leading dermatologists from around the world. A portable version of the technology is currently under development. Clinical trials are under way at St-Etienne University hospital to establish the medical interest of this imaging technology in dermatology.

## ACKNOWLEDGEMENTS

The authors thank the "Ministère de l'Enseignement Supérieur et de la Recherche (MESR)" and the "Banque Publique d'Investissement France (BPI France)" for financial supports.

## REFERENCES

[1] Bickers, D.R. et al., "The burden of skin diseases in 2004: a joint project of the American Academy of Dermatology Association and the Society for Investigative Dermatology," J. Am. Acad Dermatol 55, 490 (2006).
[2] Apalla Z. et al., "Epidemiological trends in skin cancer," Dermatol. Pract. Concept 7, 1-6 (2017).
[3] Mc Carthy M., "US melanoma prevalence has doubled over past 30 years," BMJ 350, h3074 (2015).
[4] Fuller S. R. et al., "Digital dermoscopic monitoring of atypical nevi in patients at risk for melanoma," Dermatol. Surg. 33, 1198 (2007).
[5] Kittler H., Pehamberger H., Wolff K., Binder M., "Diagnostic accuracy of dermoscopy," Lancet Oncol. 3,159 (2002).
[6] Woltsche N., Schwab C., Deinlein T., et al., "Dermoscopy in the era of dermato-oncology: from bed to bench side and retour," Expert Rev. Anticancer Ther. 16, 531-41 (2016).
[7] Glazer A.M., Rigel D.S., Winkelmann R.R., Farberg A.S., "Clinical Diagnosis of Skin Cancer: Enhancing Inspection and Early Recognition," Dermatol. Clin. 35, 409-16 (2017).
[8] Kollias N., Stamatas G.N. "Optical non-invasive approaches to diagnosis of skin diseases," J. Investig. Dermatol. Symp. Proc. 7, 64-75 (2002).
[9] Fink C., Haenssle H.A., "Non-invasive tools for the diagnosis of cutaneous melanoma," Skin Res. Technol. 23, 261-271 (2017).
[10] Pellacani G., Guitera P., Longo C., Avramidis M., Seidenari S., Menzies S., "The impact of in vivo reflectance confocal microscopy for the diagnostic accuracy of melanoma and equivocal melanocytic lesions," J. Invest. Dermatol. 127, 2759-2765 (2007).
[11] Guitera P., Pellacani G., Crotty K.A., Scolyer R.A., Li L.X., Bassoli S., Vinceti M., Rabinovitz H., Longo C., Menzies S.W., "The impact of in vivo reflectance confocal microscopy on the diagnostic accuracy of lentigo maligna and equivocal pigmented and nonpigmented macules of the face," J. Invest. Dermatol. 130, 2080-2091 (2010).
[12] Alarcon I., Carrera C., Palou J., et al., "Impact of in vivo reflectance confocal microscopy on the number needed to treat melanoma in doubtful lesions," Br. J. Dermatol. 170, 802-808 (2014).
[13] Champin J., Perrot J.L., Cinotti E., Labeille B., Douchet C., Parrau G., Cambazard F., Seguin P., Alix T. "In vivo reflectance confocal microscopy to optimize the spaghetti technique for defining surgical margins of lentigo maligna," Dermatol. Surg. Mar. 40, 247-256 (2014).
[14] Zipfel W.R. et al. "Nonlinear magic: multiphoton microscopy in the biosciences," Nat. Biotechnol. 21, 1369-1377 (2003).
[15] Koenig K. et al., "High-resolution multiphoton tomography of human skin with subcellular spatial resolution and picosecond time resolution," J. Biomed. Opt. 8, 432-439 (2003).
[16] Huang D., Swanson E.A., Lin C.P., Schuman J.S., Stinson W.G., Chang W., Hee M.R., Flotte T., Gregory K., Puliafito C.A., Fujimoto J.G., "Optical coherence tomography," Science 254, 1178-1181 (1991).
[17] Welzel J., "Optical coherence tomography of the human skin," J. Am. Acad. Dermatol. 37, 958-963 (1997).
[18] Levine A., Wang K., Markowitz O., "Optical Coherence Tomography in the Diagnosis of Skin Cancer," Dermatol. Clinics 35, 465-488 (2017).
[19] Welzel, J. "Optical Coherence Tomography in Dermatology: A Review," Skin Res. Technol. 7, 1-9 (2001).
[20] Boone M.A., Norrenberg S., Jemec G.B., Del Marmol V. "Imaging of basal cell carcinoma by high-definition optical coherence tomography: histomorphological correlation. A pilot study," Br. J. Dermatol. 167, 856-864 (2012).
[21] Coleman A.J., Richardson T.J., Orchard G., Uddin A., Choi M.J., Lacy K.E., "Histological Correlates of Optical Coherence Tomography in Non-Melanoma Skin Cancer," Skin Res. Technol. 19, 10-19 (2013).

[22] Ulrich M., Von Braunmuehl T., Kurzen H., Dirschka T., Kellner C., Sattler E., Berking C., Welzel J., Reinhold U., "The Sensitivity and Specificity of Optical Coherence Tomography for the Assisted Diagnosis of Nonpigmented Basal Cell Carcinoma: An Observational Study," Br. J. Dermatol. 173, 428-435 (2015).
[23] Podoleanu A.Gh. "Optical coherence tomography," Journal of Microscopy 247, 209-219 (2012).
[24] Drexler W., Morgner U., Kärtner F., Pitris C., Boppart S., Li X., Ippen E., Fujimoto J.G., "In vivo ultrahigh-resolution optical coherence tomography," Opt. Lett. 24, 1221-1223 (1999).
[25] Povazay B., Bizheva K., Unterhuber A., Hermann B., Sattmann H., Fercher A.F., Drexler W, et al. Opt. Lett. 27, 1800 (2002).
[26] Wang Y., Zhao Y., Nelson J.S., Chen Z., Windeler R.S., "Ultrahigh-resolution optical coherence tomography by broadband continuum generation from a photonic crystal fiber," Opt. Lett. 28, 182 (2003).
[27] Choma M.A., Sarunic M.V., Yang C., Izatt J.A., "Sensitivity advantage of swept source and Fourier domain optical coherence tomography," Opt. Express 11, 2183-2189 (2003).
[28] Ding Z., Ren H., Zhao Y., Nelson J.S., Chen Z., "High-resolution optical coherence tomography over a large depth range with an axicon lens," Opt. Lett. 27, 243-245 (2002).
[29] Leitgeb R.A., Villiger M., Bachmann A.H., Steinmann L., Lasser T., "Extended focus depth for Fourier domain optical coherence microscopy," Opt. Lett. 31, 2450-2452 (2006).
[30] Lee K.S., Rolland J.P. "Bessel beam spectral-domain high-resolution optical coherence tomography with micro-optic axicon providing extended focusing range," Opt. Lett. 33, 1696-1698 (2008).
[31] Mo J., de Groot M., de Boer J.F., "Focus-extension by depth-encoded synthetic aperture in Optical Coherence Tomography," Opt. Express 21, 10048-10061 (2013).
[32] Ralston T.S., Marks D.L., Carney P.S., Boppart S.A., "Interferometric synthetic aperture microscopy," Nat. Phys. 3, 129-134 (2007).
[33] Yu L., Rao B., Zhang J., Su J., Wang Q., Guo S., Chen Z., "Improved lateral resolution in optical coherence tomography by digital focusing using two- dimensional numerical diffraction method," Opt. Express 15, 7634-7641 (2007).
[34] Grebenyuk A., Federici A., Ryabukho V., Dubois A., "Numerically focused full-field swept-source optical coherence microscopy with low spatial coherence illumination," Appl. Opt. 53, 1697-1708 (2014).
[35] Rolland J.P., Meemon P., Murali S., Thompson K.P., Lee K.S., "Gabor-based fusion technique for Optical Coherence Microscopy," Opt. Express 18, 3632-3642 (2010).
[36] Holmes J., Hattersley S., "Image blending and speckle noise reduction in multi-beam OCT," in Optical Coherence Tomography and Coherence Domain Optical Methods in Biomedicine XIII, Proc. of SPIE 7168, 71681N (2009).
[37] Schmitt J.M., Lee S.L., Yung K.M., "An optical coherence microscope with enhanced resolving power in thick tissue," Opt. Commun. 142, 203-207 (1997).
[38] Qi B., Himmer P.A., Gordon M.L., Yang V.X.D., Dickensheets D.L., Vitkin I.A., "Dynamic focus control in high-speed optical coherence tomography based on a microelectromechanical mirror," Opt. Comm. 234, 443-448 (2004).
[39] Yang V.X.D., Munce N., Pekar J., Gordon M.L., Lo S., Marcon N.E., Wilson B.C., Vitkin I.A., "Micromachined array tip for multifocus fiber-based optical coherence tomography," Opt. Lett. 29, 1754-1756 (2004).
[40] Dubois, A., Vabre, L., Boccara, A.C., Beaurepaire, E., "High-resolution full-field optical coherence tomography with a Linnik microscope," Appl. Opt. 41, 805-812 (2002).
[41] Vabre, L., Dubois, A., Boccara, A.C., "Thermal-light full-field optical coherence tomography," Opt. Lett. 27, 530-532 (2002).
[42] Dubois, A., Grieve, K., Moneron, G., Lecaque, R., Vabre, L., Boccara, A.C., "Ultrahigh-resolution full-field optical coherence tomography," Appl. Opt. 43, 2874-2882 (2004).
[43] Dubois, A., Moneron, G., Grieve, K., Boccara, A.C., "Three-dimensional cellular-level imaging using full-field optical coherence tomography," Phys. Med. Biol. 49, 1227-1234 (2004).
[44] Dubois, A., [Handbook of full-field optical coherence microscopy, technology and applications], Pan Stanford Publishing, Singapore, 1-51 (2016)
[45] Dubois A., "Focus defect and dispersion mismatch in full-field optical coherence microscopy," Appl. Opt. 57, 142-150 (2017).
[46] Chen Y., Huang S.W., Aguirre A.D., Fujimoto J.G., "High-resolution line-scanning optical coherence microscopy," Opt. Lett. 32, 1971-1973 (2007).

[47]